\documentstyle[aps,prb,multicol,amssymb,epsf]{revtex}

\begin{document}
\draft
\title{
Dynamic Melting and Decoupling of the Vortex Lattice \\
in Layered Superconductors}
\author{Stefan Scheidl$^{1,2}$ and Valerii M. Vinokur$^2$}

\address{$^1$Institut f\"ur Theoretische Physik, Universit\"at zu
  K\"oln, Z\"ulpicher Str. 77, D-50937 K\"oln, Germany\\ $^2$Materials
  Science Division, Argonne National Laboratory, Argonne, IL 60439}

\date{\today}
\maketitle
\begin{abstract}
  The dynamic phase diagram of vortex lattices driven in disorder is
  calculated in two and three dimensions. A modified Lindemann
  criterion for the fluctuations of the distance of neighboring
  vortices is used, which unifies previous analytic approaches to the
  equilibrium and non-equilibrium phase transitions. The temperature
  shifts of the dynamic melting and decoupling transitions are found
  to scale inversely proportional to large driving currents. A
  comparison with two-dimensional simulations shows that this
  phenomenological approach can provide quantitative estimate for the
  location of these transitions.
\end{abstract}


\begin{multicols}{2}

\section{Introduction}
\label{sec.intro}

The prediction of nonequilibrium phase transitions in driven vortex
lattices \cite{KV94} has triggered an extensive theoretical study of
periodic structures moving through a random environment, using both
analytical\cite{BF95,GL96,CBFM96,BMR97,SV97.disl,BMR97.hv,GL97.hv,SV97.hv}
and numerical\cite{MSZ96,FMM96,Ryu+96,SJ97,OSZ97} approaches. The
striking experiments on vortex
transport\cite{bhatt93,kwokpl,yaron,kap,bhat96,Mar+97,Par+97} that
motivated and supported theoretical efforts, provided convincing
evidence for a genuine nonequilibrium phase transition between
different driven states of the vortex lattice. Investigations of the
$I-V$ curves in Nb$_2$Se samples and MoGe films
\cite{bhatt93,kap,bhat96} revealed regimes of plastic and elastic flow
of the vortex lattice separated by a characteristic current.  Neutron
scattering experiments \cite{yaron} on the driven vortex state in
Nb$_2$Se demonstrated a sharp increase in the density of structural
defects of the vortex lattice in a certain current interval just above
the depinning transition. At a larger characteristic current the
defect density dropped significantly, suggesting a reordering of the
vortex system.

To understand the above experiments it is crucial to determine how
quenched disorder affects the structure of the driven vortex
system. The central idea of Ref. \CITE{KV94} was the suggestion that
disorder, being frozen in the laboratory frame, appears as a
temporarily fluctuating force in the frame of the moving vortices and
leads to an increased effective temperature of the vortex system. This
effective temperature strongly depends on the drift velocity. Thus an
equilibrium phase transition of the pure system (like melting) should
have a counterpart in the system driven in a disordered environment
and could be triggered by changing the drive at constant true
temperature. For increasing drift velocities, disorder should be more
and more washed out and the temperature of the dynamic transition
should approach the transition temperature of the pure system from
below.

One related question of principal interest is to what extent
disorder-induced features of the equilibrium phase diagram could be
eliminated by applying a driving current. The most prominent example
for such a feature is the first-order melting transition of the vortex
lattice, which is experimentally found to end at a critical
point.\cite{Zeldov} For decreasing strength of disorder the location
of the transition moves to higher temperatures and the end point
wanders to larger fields.\cite{Khay+97} Therefore it seems plausible
that with increasing sample purity this critical point continues to be
shifted to larger fields and the theoretically expected phase diagram,
which does not display a critical point, is recovered. Since it is
practically impossible to obtain pinning-free samples, one might hope
to observe the phase diagram of the ideal pure system by increasing a
driving force in a given impure sample. The discussion of this
scenario is one of the goals of the present paper.

Already in the pure case one has to face the problem of how to capture
the melting transition theoretically. Here a phenomenological approach
based on the Lindemann criterion has been used
successfully.\cite{NS89,HPS89,ehB89,GK91,BGLN96} This criterion states
that the static lattice melts as the mean squared thermal displacement
of a vortex line becomes equal to a certain fraction of the lattice
spacing, $\langle {\bf u}^2({\bf r},t)\rangle \simeq c_L^2 a_0^2$. The
number $c_L$ is called the Lindemann number and is usually of order
unity. $a_0$ is the vortex spacing in the direction perpendicular to
the magnetic field. In this conventional form the Lindemann criterion
would suggest that in two dimensions (where $\langle {\bf u}^2({\bf
r},t)\rangle = \infty$ for all finite temperatures) crystals would
always be unstable to thermal fluctuations. Indeed, the long-range
translational order is lost, but a quasi-long-range translational
order and the topological order persist at low temperatures and vanish
only above a finite melting temperature. Since disorder in general
reduces long-range order to quasi-long-range order between two an four
dimensions,\cite{Nat90,GL94} it is necessary to modify the
phenomenological criterion for a detection of the more subtle loss of
topological order. This is achieved by using a slightly modified
criterion for the relative displacement of two neighbored vortices,
\begin{equation} 
w({\bf b})= {\langle [{\bf u}({\bf r}+{\bf b},t)- {\bf u}({\bf
r},t)]^2\rangle } ~.
\label{def.w} 
\end{equation} 
Here ${\bf b}$ is a basis vector of the perfect lattice that separates
the undisplaced vortices. If one naively thinks of the lattice as
being built by vortices connected by springs, then $w$ is a measure
for the typical stress of a bond (``spring'') connecting neighboring
vortices. Such bonds can be expected to ``break'' for
\begin{equation}
w({\bf b}) \approx c_{\rm L}^2 a_0^2 ~.
\label{Lindemann}
\end{equation}
In a layered superconductor, where the vortex lines are actually
composed of point-like vortices, not only the melting transition but
also a decoupling transition\cite{GK91} can exist. The evaluation of
bond widths $w$ for different orientations (in-plane of out-of-plane)
provides additional insight into the anisotropy of vortex fluctuations
and the location of these transitions.

The extension of the Lindemann criterion on disorder-dominated systems
\cite{LV95} enabled the description of disorder-induced static
transitions between elastic and plastic glasses. In view of the loss
of topological order the so-called ``entanglement'' transition from
the Bragg glass to the vortex glass is equivalent to the melting of
the pure system. Using the very same Lindemann criterion
(\ref{Lindemann}) with bond widths averaged over thermal and
disorder-induced fluctuations, the location of this entanglement
transition has been determined
recently.\cite{KNH95,EN96,Vin,Kier96,dsF97,Giamarchi,KV97} The
validity of this criterion has even been derived within a
self-consistent variational approach.\cite{KNH95}

For driven vortex systems disorder-induced displacements were
calculated in Ref. \CITE{KV94} within a naive large velocity expansion
approach for the two-dimensional vortex lattice, and the
nonequilibrium melting line was found for the case of strong
disorder. Based on scaling arguments Balents and Fisher extended the
concept of nonequilibrium phase transitions over to charge density
wave systems.\cite{BF95} Giamarchi and Le Doussal
\cite{GL96} focused on the structure of the {\it high velocity} driven
vortex phase and noticed that it retains some glassy features of its
static counterpart. The investigations that followed
\cite{CBFM96,BMR97,BMR97.hv,GL97.hv,SV97.hv} revealed further
fundamental features of the driven phase, in particular, the fact that
in a coarse-grained description (i.e., on a large enough spatial
scales) the main effect of disorder on the driven periodic structure
can be described as a {\it random force}.

Most of these recent studies have been focused on the large-scale
properties of the system in the elastic approximation. Based on the
properties of the topologically constrained system it is in principle
possible to include topological defects like dislocations and to study
the topological stability of the lattice systematically. In practice
this approach is difficult to realize, in particular when the
transition appears as collective effect of the dislocations, as in the
example of three-dimensional melting. Encouraged by the success of the
Lindemann criterion in capturing the location of the phase transition
in the pure case as well as in the static disordered case, we extend
this approach in this paper to the location of nonequilibrium
transitions of the vortex system.

The organization of this paper is as follows: in Sec. \ref{sec.pert}
we outline the perturbative approach for vortex lattices driven
through disorder and specify the generalized form of Lindemann
criterion. In Sec. \ref{sec.2D} we evaluate the Lindemann criterion
for two-dimensional systems and perform a quantitative comparison of
the resulting phase diagram with numerical simulations. In
Sec. \ref{sec.3D} we derive consequences for the dynamic phase diagram
of three-dimensional systems with a dynamic melting and decoupling
transition in \ref{sec.conc}.

\section{Perturbative approach}
\label{sec.pert}

We consider a vortex lattice in a steady driven state with velocity
${\bf v}$. Vortices are labeled by their ideal position ${\bf r}$ in
the comoving frame, whereas their actual position in the laboratory
frame is ${\bf R}({\bf r},t)={\bf r} + {\bf v} t + {\bf u}({\bf
r},t)$. The dynamics of the system is governed by the conventional
equation for the vortex displacement field ${\bf u} \equiv {\bf
u}({\bf r},t)$:
\begin{equation}
\eta \dot {\bf u}=c {\bbox \nabla}^2{\bf u} +
{\bf f}^{\rm pin}({\bf R})+ {\bf F}-\eta {\bf v} + {\bbox \zeta}
\label{eq.mo}
\end{equation}
where $c$ represents the elastic constants (specified below for the
vortex lattice) and ${\bbox \nabla}$ the lattice gradient. The thermal
noise ${\bbox \zeta}$ couples the vortices to a heat bath of
temperature $T$. The pinning force density ${\bf f}^{\rm pin}({\bf
r},t) =- {\bbox \nabla}V[{\bf R}({\bf r},t)]$ is related to the
pinning potential $V$, which is supposed to be Gaussian distributed
with a second moment
\begin{mathletters} \label{Phi}
\begin{eqnarray}
\overline
{V({\bf k}) V({\bf k}')}& =&
\Delta ({\bf k})\delta({\bf k}+{\bf k}'),
\\
\Delta ({\bf k}) &=& \Delta_0 e^{- \frac 12 k_\perp^2 \xi^2}.
\end{eqnarray}
\end{mathletters} 
The correlations decay on the scale of the coherence length $\xi$. We
define $k_\perp^2=k_x^2+k_y^2$ as the vector component perpendicular
to the magnetic field. Eq. (\ref{eq.mo}) refers to point-like vortices
in a single layer ($D=2$) or in a layered superconductor ($D=3$).

In the absence of disorder the response of the vortex lattice is
different for longitudinal (L) and transverse (T) modes. Since the
vortex lattice is almost incompressible, the longitudinal modes do not
significantly contribute to many physical properties. However, as we
see later on, in a vortex lattice driven through disorder these modes
are important and have to be retained. The response function $G$ is
determined by the elastic constants for compression $c_{11}$, shear
$c_{66}$ and tilt $c_{44}$:
\begin{mathletters} \label{response}
\begin{eqnarray}
G_{\alpha\beta}({\bf q},\omega)&=&
\sum_p
G^p({\bf q},\omega)  P^p_{\alpha\beta} ({\bf q}) ~,
\\
G^p({\bf q},\omega) &=&  [ -i \eta \omega
+ c_p q_\perp^2 + c_{44} q_z^2 ]^{-1} ~,
\\
P^T_{\alpha\beta} ({\bf q})&=&
\delta_{\alpha \beta}- \frac {q_\alpha q_\beta}{q_\perp^2} ~,
\quad P^L_{\alpha\beta} ({\bf q})=
\frac {q_\alpha q_\beta}{q_\perp^2},
\end{eqnarray}
\end{mathletters}
where $p=L,T$ stands for a polarizations with $c_p=c_{11},c_{66}$
respectively, and $q_\perp^2=q_x^2+q_y^2$. Wave vectors ${\bf q}$ are
restricted to the first Brillouin zone (BZ) of the ideal lattice in
contrast to ${\bf k}$.

To calculate the fluctuations $w$ of the distance of neighboring
vortices we treat disorder on the lowest level of perturbation theory
by approximating ${\bf f}({\bf R})={\bf f} ({\bf r} + {\bf v}t + {\bf
u}) \approx {\bf f} ({\bf r} + {\bf v}t) $, which is justified as long
as $w({\bf b}) \lesssim \xi^2$. This condition is satisfied in the
whole range of interest $w({\bf b}) \leq c_L^2 a_0^2$ if $c_L a_0
\lesssim \xi$. This restricts the validity of our approach to high
magnetic fields not too much below the upper critical field.

In this approximation the pinning force can be considered as
``external'' force that does not depend on the response of the vortex
lattice. The combination ${\bf f} ({\bf r},t)= {\bf f}^{\rm pin} ({\bf
r}+ {\bf v} t) + {\bbox \zeta}({\bf r},t)$ of this pinning force and
the thermal noise has Gaussian correlations
\begin{mathletters} 
\label{corr.f}
\begin{eqnarray}
\overline{\langle f_\alpha({\bf q},\omega) f_\beta({\bf q}',\omega') 
\rangle} &=&
\Psi_{\alpha \beta} ({\bf q},\omega)
\delta({\bf q}+{\bf q}') \delta(\omega + \omega'),
\\
\Psi_{\alpha \beta}({\bf q},\omega) &=&
\vartheta \delta_{\alpha \beta}
+ \sum_{\bf Q} k_\alpha k_\beta \Delta ({\bf k})
\delta(\omega + {\bf v} \cdot {\bf k}),
\nonumber \\
\vartheta &=& 2 \eta T.
\nonumber 
\end{eqnarray}
\end{mathletters}
The averaging includes the randomness of disorder and that of thermal
noise.  We denote ${\bf k}={\bf Q}+{\bf q}$ with a reciprocal lattice
vector (RLV) ${\bf Q}$ (being perpendicular to the magnetic field) and
a vector ${\bf q}$ within the BZ.

The displacements in response to the total external force ${\bf f}$
have correlations
\begin{mathletters} 
\label{corr.u}
\begin{eqnarray}
\overline{u_\alpha({\bf q},\omega)
  u_\beta({\bf q}',\omega')} &=& C_{\alpha \beta}({\bf q},\omega)
\delta({\bf q}+{\bf q}') \delta(\omega + \omega')
\\
C_{\alpha \beta}({\bf q},\omega)
&=&G_{\alpha \gamma}({\bf q},\omega)
\Psi_{\gamma \delta} ({\bf q},\omega)
G_{\beta \delta}(-{\bf q},-\omega).
\nonumber 
\end{eqnarray}
\end{mathletters} 
From Eq. (\ref{corr.u}) we can immediately calculate the bond
fluctuations
\begin{equation}
\label{w}
w ({\bf b}) \approx \int_{\omega \bf q}
\frac 12 ({\bf q} \cdot {\bf b})^2
C_{\alpha \alpha} ({\bf q},\omega) ~.
\end{equation}
The thermal and pinning contributions to the force correlator
(\ref{corr.f}) generate two corresponding contributions to the bond
width, $w=w^{\rm th} + w^{\rm pin}$, which can be considered as
approximately independent. The {\it thermal} contribution has been
studied extensively in the past in order to determine the melting
transition of the vortex lattice in the absence of impurities. Here we
focus our attention on the contribution due to pinning, which is
\begin{equation}
\label{w.pin}
w^{\rm pin} ({\bf b}) \approx \sum_{p,\bf Q} \int_{\bf q}
\frac 12 
{({\bf q} \cdot {\bf b})^2}
\Gamma^p ({\bf k}) \Delta({\bf k}) ~,
\end{equation}
where $\int_{\bf q} = \int \frac {d^D q}{(2 \pi)^D}$ and we abbreviate
\begin{mathletters} \label{def.Gamma}
\begin{eqnarray}
\Gamma^T ({\bf k}) &\equiv& 
\frac {({\bf k}_\perp \wedge {\bf q}_\perp)^2}
{ q_\perp^2} 
\frac 1 
{\eta^2 ({\bf v} \cdot {\bf k})^2 + (c_{66} q_\perp^2 + c_{44}
q_z^2)^2} ~,
\\
\Gamma^L  ({\bf k}) &\equiv& 
\frac {({\bf k}_\perp \cdot {\bf q}_\perp)^2 }{ q_\perp^2}
\frac 1 
{\eta^2 ({\bf v} \cdot {\bf k})^2 + (c_{11} q_\perp^2 + c_{44}
q_z^2)^2} ~.
\end{eqnarray}
\end{mathletters} 

Before we proceed to a detailed evaluation of Eq. (\ref{w.pin}) in the
following sections, we cite the result of lowest order perturbation
theory for the macroscopic friction force ${\bf F}^{\rm fr}$ arising
from the collective summation of the microscopic pinning forces. In
order to achieve a drift velocity ${\bf v}$ one has to apply a driving
force ${\bf F}({\bf v})= \eta {\bf v} + {\bf F}^{\rm fr}({\bf v})$
with\cite{SV97.hv}
\begin{mathletters} \label{F.fr}
\begin{eqnarray}
F^{\rm fr}_\alpha&=& \int_{\bf k}
i k_\alpha k_\beta k_\gamma \Delta({\bf k})
G_{\beta \gamma}({\bf k},-{\bf v} \cdot {\bf k})
\\
&=& \sum_{p, \bf Q} \int_{\bf q}
{k_\alpha \eta ({\bf v} \cdot {\bf k} )}
\Gamma^p ({\bf k}) \Delta({\bf k}) ~,
\end{eqnarray}
\end{mathletters} 
which coincides for $D=2$ with the early result of Schmid and
Hauger.\cite{SH73}

The disordered vortex lattice is characterized by two elementary
length scales: the disorder correlation length $\xi$ and the ``vortex
spacing'' $a_0 \equiv \sqrt{\Phi_0/B}$ ($B$ is the magnetic induction
and $\Phi_0$ the flux quantum).  In the triangular lattice the actual
distance between neighboring vortices in a direction perpendicular to
the magnetic field is $a=\sqrt{2/\sqrt{3}} a_0$.  We assume $c_L a_0
\lesssim \xi \lesssim a_0$ which is realistic for high temperature
superconductors at large fields. For the evaluation of the main
formulae (\ref{w.pin}) and (\ref{F.fr}) we will retain the reciprocal
lattice structure in ${\bf Q}$, but we approximate the BZ as spherical
cylinder with bounds ${\bf q}_\perp ^2 \leq {q_\perp^*}^2 \equiv 4 \pi
/a_0^2$ and $|q_z| \leq {q_z^*} \equiv \pi/d$ with the layer spacing
$d$. This approximation preserves the area of the BZ.  The nonlocality
of $c_{66}$ will be neglected. We choose the $x$-axis as the direction
of the velocity, ${\bf v}= v \hat {\bf x}$, and suppose the vortex
lattice to move along one main direction of the hexagonal lattice,
which are the directions of minimum energy dissipation.\cite{SH73}

\section{Evaluation in $D=2$}
\label{sec.2D}

The general expressions for bond widths $w$ and friction force ${\bf
F}^{\rm fr}$ are immediately specialized to two dimensions by setting
$q_z=0$. They are evaluated here in order to compare the location of
the dynamic melting transition according to the Lindemann criterion
with numerical simulations.

\subsection{Thermal bond width}

To start with we consider the contribution of thermal fluctuations to
the bond width. By a comparison with the Kosterlitz-Thouless melting
theory \cite{KT73} we are able to fix the Lindemann number $c_L$.

Thermal fluctuations lead to a displacement correlation
\begin{equation}
C^{\rm th}_{\alpha \alpha} ({\bf q},\omega) =
\frac {\vartheta}{\eta^2 \omega^2 + c_{66}^2 q^4} ~,
\label{2D.thermal.C}
\end{equation}
which implies a bond width
\begin{equation}
w^{\rm th}({\bf b}) \approx \int_{\omega \bf q}
\frac 12 ({\bf q}\cdot {\bf b})^2
C^{\rm th}_{\alpha \alpha} ({\bf q},\omega)
\approx b^2 \frac {T}{4 a_0^2  c_{66}} ~.
\label{2D.thermal.w}
\end{equation}
Here the contribution of longitudinal displacements can be safely
neglected since typically $c_{11} \gg c_{66}$. In $D=2$ there is only
one bond length $b=a$. The actual melting temperature within
Kosterlitz-Thouless theory \cite{KT73} (neglecting renormalization
effects) is given by
\begin{equation}
\label{T_KT}
T_{\rm KT} \approx \frac{a_0^2 c_{66}} {4 \pi} ~.
\end{equation} 
Inserting this temperature into Eq. (\ref{2D.thermal.w}) one can
estimate the Lindemann parameter
\begin{equation}
  c_{\rm L}^2 \approx \frac{1}{6 \sqrt 3 \pi} \approx 0.03 ~.
\label{c_L}
\end{equation}
This value is in good agreement with typical values that empirically
describe the melting transition in high temperature
superconductors. We stick to this value in all quantitative
considerations below.

As soon as we have determined the pinning bond widths we can estimate
the location of the transition where bonds ${\bf b}$ become unstable
according to the Lindemann criterion $w^{\rm th} +w^{\rm pin}=c_L^2
a_0^2$. The transition temperature will be determined by
\begin{equation}
T_{\rm c}=T_{\rm KT}- \frac {4 a^2 c_{66}} {a_0^2} w^{\rm pin} 
\end{equation}
as a function of the bond orientation and of velocity.

\subsection{Pinning bond width}

The evaluation of the disorder contributions (\ref{w.pin}) to the bond
width is more subtle than the calculation of the friction force. In
particular, since the velocity selects a particular direction, which
is chosen parallel to the $x$-axis (called $x$-bonds), we have to
distinguish bonds pointing parallel to the velocity and bonds that
enclose an angle of 60 degrees with the $x$-axis (called $y$-bonds).
Both types of bonds have a length $|{\bf b}|=a$. Due to the invariance
of Eq. (\ref{w.pin}) under a reflection of components of ${\bf k}$,
one can parameterize the bond width by two coefficients,
\begin{equation}
w^{\rm pin} ({\bf b}) =
w^{\rm pin}_x b_x^2/b^2 + w^{\rm pin}_y b_y^2/b^2  ~.
\label{w.matrix}
\end{equation}
These coefficients $w^{\rm pin}_x$ and $w^{\rm pin}_y$ can be
calculated analytically for small and large velocities:
\begin{mathletters} \label{w.2D}
\begin{eqnarray}
w^{\rm pin}_x &\approx&
\left\{
\begin{array}{l}
\frac{a^2 \Delta_0}{\sqrt{8 \pi} a_0 \xi^3 \eta^2 v^2} ~,
\\
\frac{a^2 a_0^2 \Delta_0} {32 \pi^2 \xi^4 c_{66}^2}
\left(1+\frac {3 \sqrt{2 \pi} \xi}{2a_0}\right)
\ln \left( \frac{4 \pi \xi c_{66}}{a_0^2 \eta v} \right) ~,
\end{array}
\right.
\label{w_xx}
\\
w^{\rm pin}_y &\approx&
\left\{
\begin{array}{l}
\frac{a^2 \Delta_0}{4\sqrt{2 \pi} \xi^3 \eta v c_{11}} ~,
\\
\frac{a^2 a_0^2 \Delta_0} {32 \pi^2 \xi^4 c_{66}^2}
\left(1+\frac {\sqrt{2 \pi} \xi}{2a_0}\right)
\ln \left( \frac{4 \pi \xi c_{66}}{a_0^2 \eta v} \right) ~.
\end{array}
\right.
\label{w_yy}
\end{eqnarray}
\end{mathletters}
Upper and lower expressions hold for $v \gg {c_{11}} / {a_0 \eta} $
and $v \ll {\xi c_{66}} /{a_0^2 \eta} $ respectively.
 
In the limit of large velocities the contributions to leading order in
$v$ and in small $\xi /a_0$ come only from RLV with $Q_x=0$, since
then the denominator of the response function becomes small, $\eta^2
v^2 k_x^2 + c_p^2 q^4 =\eta^2 v^2 q_x^2 + c_p^2 q^4$, and gives
largest weight to small $q_x$. In the case of $w^{\rm pin}_x$ the
elastic interaction is negligible above a characteristic velocity $v
\sim c_p/a_0 \eta $. Then longitudinal and transverse modes are {\em
equally} important and give contributions $\sim v^{-2}$. For $w^{\rm
pin}_x$ vortices therefore respond as if they were independent
particles. In the case of $w^{\rm pin}_y$ there is a {\em qualitative}
difference between longitudinal and transverse modes. Transverse modes
again give only contributions $\sim v^{-2}$, whereas now longitudinal
modes give dominating contributions of order $\sim v^{-1}$.  Thus at
large velocities $y$-bonds are subject to much stronger fluctuations
than $x$-bonds.

For small velocities the elastic interaction dominates the response of
vortices and for $c_{11} \gg c_{66}$ longitudinal modes are in general
negligible compared to the transverse modes. The small velocity regime
is reached when in the denominator $\eta^2 v^2 k_x^2 + c_p^2 q^4$ of
the response functions the first term is typically small compared to
the second one, i.e. below a velocity $v \sim \xi c_p/a_0^2 \eta $. To
the leading order in small $v$ and $\xi/a_0$, RLV of all directions
contribute equally the the bond widths, which diverge $\sim \ln
(1/v)$. The anisotropy does {\em not} vanish at small velocities, the
prefactor of the logarithmic divergence depends (in subleading order
in $\xi/a_0$, arising from contributions of RLV with $Q_x=0$) on the
bond orientation. In contrast to the high velocity regime, $x$-bonds
have stronger fluctuations than $y$-bonds at small velocities.

These results will be discussed in more detail later in section
\ref{sec.disc.2D}.

\subsection{Friction force}

In most experiments and simulations the driving force rather than the
velocity is the parameter controlling the drift. Therefore we wish to
express the bond widths as a function of the force and to evaluate the
transport characteristics ${\bf v}({\bf F})$ from the friction force
(\ref{F.fr}). Since we assume that vortices drift along a basic
lattice direction, the friction force is also parallel to velocity,
${\bf F}^{\rm fr} \parallel {\bf v} \parallel {\bf
x}$. Eq. (\ref{F.fr}) can be evaluated analytically in the limits of
large and small velocities. In both cases the main contributions to
the sum over RLV's come from ${\bf Q}$ with all possible
orientations. Therefore the characteristic velocity separating the
large and small velocity regimes for transverse and longitudinal modes
are given by $v \sim \xi c_{66} / a_0^2 \eta$ and $v\sim \xi c_{11}/
a_0^2 \eta$ respectively. We find
\begin{mathletters} \label{F.fr.approx}
\begin{eqnarray}
F^{\rm fr} &\approx&
\sum_{p, \bf Q} \Delta({\bf Q}) \frac {Q^2 Q_x }{ 8 \pi c_p}
\ {\rm atn} \frac {4 \pi c_p}{Q_x a_0^2 \eta v}
\\
& \approx &
\left\{
\begin{array}{ll}
\frac{\Delta_0}{\pi \xi^4 \eta v} 
& {\rm for} \ v \gg \xi c_{11}/a_0^2 \eta ~,
\\
\frac{ 3 a_0^2 \Delta_0}{2 (2 \pi)^{5/2} \xi^5 c_{66}}
& {\rm for} \ v \ll \xi c_{66}/a_0^2 \eta ~.
\end{array}
\right.
\end{eqnarray}
\end{mathletters}
Only for large velocities $v \gg \xi c_{11}/a_0^2 \eta $ longitudinal
and transverse modes contribute equally to the friction force. At
velocities $v \ll \xi c_{11}/a_0^2 \eta $ the transverse modes dominate
since $c_{11} \gg c_{66}$. Note that the friction force enters the
large velocity regime already for $v \gg \xi c_{11}/a_0^2 \eta $,
whereas the bond widths reach their corresponding regime only for $v
\gg c_{11}/a_0 \eta $.

\subsection{Numerical evaluation}

In order to illustrate the analytic results and to demonstrate the
capability of the Lindemann approach to provide even a {\em
quantitative} estimate for the location of dynamic phase transitions,
we perform a numerical evaluation of the bond width and the friction
force from Eqs. (\ref{w.pin}) and (\ref{F.fr}). The results are then
compared with the simulation data in Ref. \CITE{KV94}.

For this purpose we specify the parameters as follows.  In
Ref. \CITE{KV94} the vortex spacing $a$ was used as length scale and
$2 d \epsilon_0 \equiv 2 d (\Phi_0/4 \pi \lambda)^2$ as energy scale
($\lambda$ is the penetration depth, $d$ is the layer thickness). The
time scale is set by $\eta=1$. From the melting temperature $T_{\rm
KT} \approx 0.007$ of the pure system we find $c_{66} \approx 0.088$
according to Eq. (\ref{T_KT}). The vortex interaction chosen in
Ref. \CITE{KV94} decays on the penetration length estimated by
$\lambda \approx a$. The compression constant is obtained from $c_{11}
\approx (16 \pi \lambda^2/a^2) c_{66} \approx 50 c_{66}$.\cite{blat94}
We furthermore identify $\xi \equiv r_p=0.2$ and refer to the data
sets with a pinning strength $A=0.006$, a number of pinning centers
$N_p=10^4$, and a number of vortices $N_v=400$. Then the disorder
strength is $\Delta_0=\gamma_U/a_0^2=(N_p/N_v a_0^2) [\pi A
(\xi/a_0)^2]^2 \approx 1.42 \cdot 10^{-5}$.

\begin{figure}
\epsfxsize=0.9 \linewidth
\epsfbox{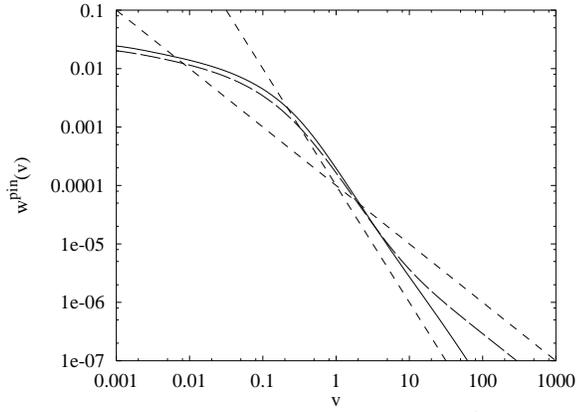}
\narrowtext
\caption{
Plot of pinning bond widths $w^{\rm pin}_x$ (full line) and $w^{\rm
pin}_y$ (long dashed line) calculated numerically from
Eq. (\ref{w.pin}) for the parameters specified in the text. The short
dashed lines represent dependences $w^{\rm pin}(v)
\sim v^{-1}, v^{-2}$, which are realized at high $v$.}
\label{fig_wpin}
\end{figure}

Fig. \ref{fig_wpin} shows the pinning bond widths for $x$-bonds (full
line) and $y$-bonds (long dashed line). The short dashed lines are a
guide to the eyes representing an asymptotic decay $\sim v^{-2}$ for
$x$-bonds and $\sim v^{-1}$ for $y$-bonds. This asymptotic regime is
reached only for $v \gtrsim 100$. At small velocities $v \lesssim
0.01$ both bond widths diverge logarithmically. For large velocities,
$y$-bonds have stronger fluctuations than $x$-bonds, as opposed to
small velocities.

\begin{figure}
\epsfxsize=0.9 \linewidth
\epsfbox{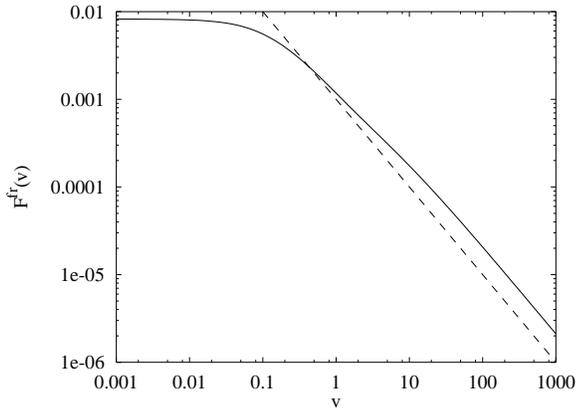}
\narrowtext
\caption{
Plot of friction force (full line) calculated numerically from
Eq. (\ref{F.fr}) for the parameters specified in the text. The dashed
line represent the dependence $F^{\rm fr}(v)
\sim v^{-1}$ at high $v$.}
\label{fig_Ffr}
\end{figure}

\begin{figure}
\epsfxsize=0.9 \linewidth
\epsfbox{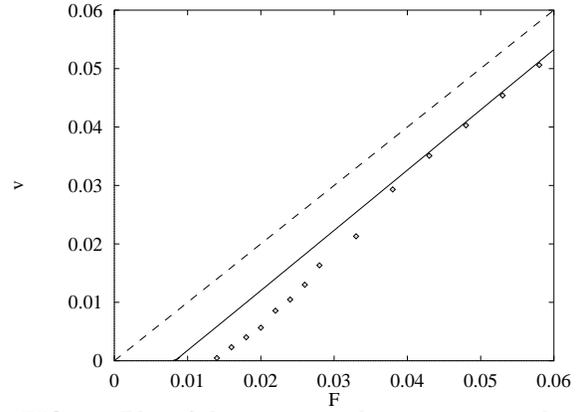}
\narrowtext
\caption{
Plot of the transport characteristic resulting from Eq. (\ref{F.fr}),
full line, and the simulation data of Ref. \protect\CITE{KV94} (dashed
line). }
\label{fig_v}
\end{figure}

The friction force obtained from Eq. (\ref{F.fr}) is shown in
Fig. \ref{fig_Ffr} by the full line. The dashed line displays a
$v^{-1}$ dependence, which is realized by $F^{\rm fr}$ for $v \gtrsim
100$. At small velocities, $v \lesssim 0.01$, $F^{\rm fr}$ saturates
at a finite value in accordance with Eq. (\ref{F.fr.approx}b). The
resulting transport characteristics $v(F)$ is shown in
Fig. \ref{fig_v}. The dashed line is the characteristics in the
absence of pinning, which is shifted to the full line by the presence
of disorder. This shift is practically constant in the small velocity
regime. The dots represent the simulation data of Ref. \CITE{KV94} for
the lowest temperature ($T=0.001$) considered there. The agreement
with the perturbative result is surprising for $F > F_c \approx 0.04$,
where vortices move coherently ($F_c$ corresponds to $f_t$ in
Ref. \CITE{KV94}). The regime of forces $F < F_c$ is beyond the
validity range of our elastic approach.

\begin{figure}
\epsfxsize=0.9 \linewidth
\epsfbox{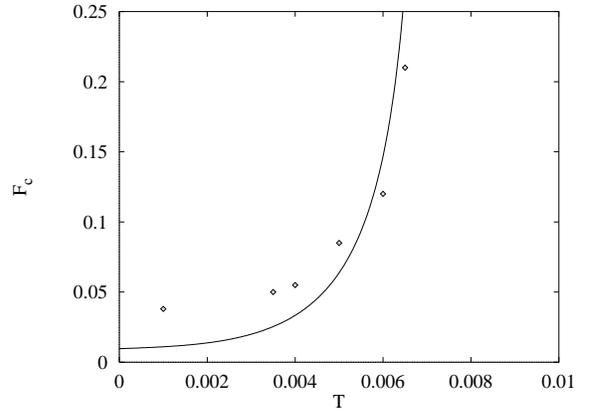}
\narrowtext
\caption{
Plot of critical force for the transition from coherent to incoherent
vortex motion. The full line is our result from the Lindemann
criterion, the dots are simulation results from
Ref. \protect\CITE{KV94}.}
\label{fig_Fc}
\end{figure}

Fig. \ref{fig_Fc} displays the resulting phase diagram, comparing the
result of the Lindemann criterion (full line) to the simulation
(dots). Above this transition line all bonds are stable, below the
transition line the Lindemann criterion breaks the breaking bonds. In
the displayed range of small velocities $x$-bonds are the least
stable. Agreement between perturbation theory and simulations is given
up to a factor of the order of unity, which is still quite favorable
in view of the conceptual simplicity of the Lindemann approach and its
sensitivity to changes e.g. of the Lindemann number. It is worthwhile
to state that no fit parameters have been used in our numerical
calculation.

\subsection{Discussion}
\label{sec.disc.2D}

At this point we rest to discuss some specific properties of our
results (\ref{w_xx}) and (\ref{w_yy}) for the bond widths.

The first property is the strong anisotropy at large velocity, where
the width of $x$-bonds scales like $\sim 1/v^2$ independent of the
elastic constants, whereas the $y$-bond width scales like $\sim 1/v
c_{11}$. This is true only for largest velocities $v \gg c_{11}/a_0
\eta $, whereas for $v \ll c_{11}/a_0 \eta $ the isotropy is
essentially restored, i.e. $y$-bonds have {\em qualitatively} the same
velocity dependence as $x$-bonds but the prefactors are still
different. At large $v$ the physical origin of the anisotropy can be
understood in a simple picture where vortices are considered as
independent particles. Vortices neighbored in $x$-direction follow the
same paths and are exposed to the same pinning forces. They experience
the same force, but with a delay time $\Delta t = a/v$. Even in the
absence of vortex interactions such purely time shifted forces give
rise to bond fluctuations of a finite width only. Therefore the
elastic interaction is irrelevant and these bond fluctuations are
independent of elastic constants and decay proportional to $(\Delta
t)^2 \propto v^{-2}$. Vortices neighbored in directions not parallel
to velocity always move on different trajectories and are exposed to
essentially uncorrelated pinning forces. In the absence of
interactions their typical relative distance would increase without
limits in the direction perpendicular to ${\bf v}$ as in a diffusion
process. In the lattice such a diffusive motion is prevented by the
vortex interactions. Since their relative distance fluctuates mainly
in $y$-direction, which is almost parallel to their distance, it is
the compression modulus rather than the shear modulus that confines
the bond fluctuations.

The anisotropy at large $v$ can be related to the anisotropy of the
Larkin domain in the driven lattice.\cite{GL96,SV97.hv} This domain is
much longer in the direction parallel to the velocity than in the
other directions, which means that relative displacements grow faster
in direction perpendicular to ${\bf v}$ than parallel to ${\bf
v}$. Our conclusion that $y$-bonds are less stable than $x$-bonds at
large velocities is in agreement with a smectic structure of the
drifting system,\cite{BMR97,BMR97.hv,GL97.hv,SV97.hv} where vortices
move in decoupled chains aligned parallel to ${\bf v}$.

As second distinct feature of the bond fluctuations, the divergence of
the bond widths at small velocities deserves some explanation. This
divergence, which is in general found for $D \leq 2$, implies that at
zero velocity and for arbitrarily weak disorder all bonds are broken
and the vortex lattice is destroyed, i.e. that the structure factor
resembles that of a liquid. This result of the Lindemann criterion
coincides surprisingly with that of more elaborate methods,
e.g. renormalization group methods,\cite{CO82,VF84} which evaluate the
large-scale correlations of the displacement field. The divergence of
the bond width in our approach arises from the integration over small
momenta ${\bf q}$, i.e. from large scale fluctuations. From a
principal point of view our perturbative treatment of disorder is not
valid at largest scales, where higher order effects can no longer be
neglected. Therefore the correctly found divergence of the bond width
has to be considered as lucky circumstance.

Although the phenomenological Lindemann criterion is able to capture
the position of a given transition, it can neither proof the existence
of a transition nor can it tell us something about the nature of a
transition. Although in 2D a phase transition can be observed in the
density of lattice defects and, as a consequence, in specific features
of the transport characteristics,\cite{bhatt93,kap} scaling arguments
\cite{BF95} and numerical calculations of the structure factor
\cite{MSZ96,SJ97} indicate that on {\em large} length scales free defects
should exist, i.e. that the topological order of the lattice is
lost. However, at present a rigorous characterization of the
large-scale properties and examination of the effects of defects is
still missing. Even if a true phase transition (like
``solid-to-fluid'') related to a qualitative change of the large
correlations might be absent in 2D case, we expect a transition (like
``liquid-to-gas'') or at least a pronounced crossover at the location
of the dynamic phase ``transition'' located by the Lindemann
criterion, which in its generalized form used here probes small scale
correlations. While these displacement fluctuations on small scales
are related to the {\em rate of generation} of (initially bound)
dislocation pairs, large scale fluctuations of the displacement field
are relevant to decide whether dislocations remain {\em bound} in
pairs or dissociate into free dislocations that destroy the
topological order.

It is instructive to draw a comparison between the bond widths, which
we examine here, and the shaking temperature $T_{\rm sh}$ introduced
in Ref. \CITE{KV94}. $T_{\rm sh}$ was defined from the correlator of
the pinning force experienced by a single particle. It is therefore
independent of elastic constants. If we compare the thermal and
disorder contributions to the bond width, the latter could be
expressed in terms of an effective ``bond shaking temperature''
\begin{equation}
T^{\rm bsh}({\bf b}) \equiv \frac {4 a_0^2 c_{66}}{b^2} w ({\bf b}) ~.
\label{def.Tbsh}
\end{equation}
This bond shaking temperature differs from $T_{\rm sh}$ in several
respects. First of all it depends on the orientation of the bonds
under consideration.  At large velocities $T_{\rm sh}$ and $T^{\rm
bsh}$ for $y$-bonds have the dependence $\sim v^{-1}$ in
common. However they differ in the prefactor, which is independent of
the elastic constants in the former case (since $T_{\rm sh}$
characterizes a single particle), but contains a prefactor $\sim
c_{66} /c_{11}$ in the latter case (since $T^{\rm bsh}$ characterizes
a relative displacement response).  $T^{\rm bsh}$ is more similar to
the coherent shaking temperature $T_{\rm sh}^{\rm coh}$ of
Ref. \CITE{KV94}, which was found to decay $\sim v^{-2}$ in agreement
with $T_{\rm bsh}$ for $x$-bonds (and even $y$-bonds in an
incompressible lattice).

Speaking about effective temperatures, the attribute ``shaking''
should not be taken too literally, since even for $v=0$ the bond width
$w^{\rm pin}$ is finite and static, i.e. it has not temporal
fluctuations. However, a snapshot of the displacements on small scales
looks like that of a system without disorder but with a temperature
increased by $T^{\rm bsh}$. Since an effective temperature cannot be
defined uniquely in a nonequilibrium situation anyway, we keep in the
following the notion of bond fluctuations. Since these fluctuations
are related to the fluctuations of the Peach-Koehler force acting on
dislocations, we believe that the bond fluctuations are an appropriate
measure for the relevance of dislocation (at least on not too large
scales, see discussion above). The order-of-magnitude agreement on the
location of the transition between the Lindemann criterion and
simulation supports this picture.

\section{Evaluation in 3D}
\label{sec.3D}

In a layered three-dimensional superconductor the physics is even more
rich than in two dimensions: besides the melting transition, where the
structural order of the vortex lattice gets lost and its elastic
moduli get renormalized to zero, an additional decoupling transition,
where the conductivity perpendicular to the layers becomes ohmic and
the effective Josephson coupling between the layers gets lost, can
occur. We evaluate the bond fluctuations for layered superconductors
in the driven disordered case and discuss the implications for the
nonequilibrium counterpart of the equilibrium melting and decoupling
transitions.

In bulk superconductors, in particular for anisotropic
high-temperature superconductors, the elastic properties of the vortex
lattice are somewhat intricate due to the dispersion of the elastic
constants (mainly of $c_{44}$ and $c_{11}$). For completeness we cite
the values\cite{HPS89,GK91}
\begin{mathletters} \label{c.3D}
\begin{eqnarray}
c_{11}&=& \frac{\epsilon_0}{a_0^2} 
\frac{\lambda^2 {q_\perp^*}^2}
{1+\gamma^2 \lambda^2 q_\perp^2 + \lambda^2 q_z^2}
\frac {1+\gamma^2 \lambda^2 q^2 }
{1+\lambda^2 q^2} ~,
\\
c_{66} &=&
\frac {\epsilon_0}{4a_0^2} ~,
\\
c_{44}&=& c_{44}^0 + c_{44}^c  ~,
\\
c_{44}^0&=& \frac {\epsilon_0}{a_0^2} 
\frac{\lambda^2 {q_\perp^*}^2} 
{1+\gamma^2 \lambda^2 q_\perp^2 + \lambda^2 q_z^2} ~,
\\
c_{44}^c&=&\frac {\epsilon_0}{2 a_0^2} 
\Bigg[ \frac 1 {\gamma^2} \ln \left(
\frac{\gamma^2 \xi^{-2}}{\lambda^{-2}+\gamma^2 {q_\perp^*}^2+q_z^2}
\right)
\nonumber \\ &&
+\frac1{\lambda^2 q_z^2}
\ln \left(1 + \frac{\lambda^2 q_z^2}{1+\lambda^2 {q_\perp^*}^2} \right) \Bigg] ~.
\end{eqnarray}
\end{mathletters}
We define ${q_\perp^*} \equiv \sqrt{4 \pi}/a_0$, ${q_z^*} \equiv \pi/d$, and
$\epsilon_0=(\Phi_0/4 \pi \lambda)^2$. The tilt modulus is composed of
a nonlocal contribution to the tilt energy and a ``single-vortex''
contribution.\cite{GK91} We are interested in the range of not too
small fields, where $\lambda \gtrsim a_0$ and an exponential decay of
the elastic constants with the vortex density can be neglected.

\subsection{Thermal bond widths}

It is straightforward to evaluate the bond fluctuations due to thermal
fluctuations and in the absence of disorder. Although purely thermal
fluctuations have already been well examined within the usual
Lindemann approach\cite{NS89,HPS89,ehB89,BGLN96}, we calculate the
bond fluctuations in order to gain additional insight into the
anisotropic nature of the fluctuations. Neglecting again the
contributions of compression modes, we find
\begin{mathletters} \label{w.th.3d}
\begin{eqnarray}
w^{\rm th}_x &\equiv& w^{\rm th}(a \hat{\bf x})  \approx 
\frac{a^2 T}{4 d a_0^2 c_{66}} \frac \beta {\sqrt{2}} 
\ {\rm atn} ( \frac{\sqrt{2}} \beta) 
\nonumber \\
& \approx & 
\left\{
\begin{array}{ll} 
\frac{a^2 T}{4 a_0^2 d c_{66}} 
& {\rm for } \quad \beta \gg 1 \\
\frac{\pi a^2 T}{8 \sqrt 2 a_0^2 d c_{66}} \beta
& {\rm for } \quad \beta \ll 1 ~,\\
\end{array}
\right.
\\
w^{\rm th}_z &\equiv& w^{\rm th}(d \hat{\bf z})  \approx 
\frac {\pi T}{24 d c_{66}}
\left[
2 \beta^2- 2 \beta^3 \ {\rm atn} \frac 1 \beta + \ln (1+ \beta^2)
\right]
\nonumber \\
& \approx& 
\left\{
\begin{array}{ll} 
\frac{\pi T}{12 dc_{66}} \ln \beta 
& {\rm for } \quad \beta \gg 1 \\
\frac{\pi T}{8 d c_{66}} \beta^2
& {\rm for } \quad \beta \ll 1 ~, \\
\end{array}
\right.
\end{eqnarray}
\end{mathletters} 
where 
\begin{equation}
\beta^2 \equiv \frac {c_{66} {q_\perp^*}^2}{c_{44} {q_z^*}^2}
=\frac{B}{B_{cr}}
\end{equation}
is the magnetic field in units of the crossover field\cite{GK91,FFH91}
\begin{equation}
B_{cr}=\frac {\pi \Phi_0 c_{44}}{4 d^2 c_{66}} ~.
\end{equation}

For the transparency of our arguments and for qualitative purposes we
have suppressed the explicit nonlocality of the elastic constants. The
nonlocality can be restored by using effective values of the
dispersive elastic constants (\ref{c.3D}) evaluated at the length
scales that give the largest contributions to the integrals.  In $D=3$
bond fluctuations are always dominated by small-scale fluctuations,
since even in the presence of disorder and for small velocities one
never encounters a divergence arising from large scales (unlike in $D
\leq 2$). Therefore we use in our following estimates elastic constants
evaluated for $q_\perp={q_\perp^*}$ and $|q_z|={q_z^*}$. In general this
approximation underestimate the stiffness of the lattice but it is
qualitatively valid as long as the dispersion (anisotropy of the
elastic constants) is not too pronounced.

Let us now locate the lines where $w^{\rm th}_\alpha = c_l^2 a_0^2$
for all bond types $\alpha = x,y,z$. These lines can be described by
$T_\alpha \equiv T_\alpha(B)$ in the $(T,B)$ plane. To be specific, we
consider in the following moderately anisotropic systems with $\gamma d
\ll \lambda $ (like for YBCO, and even BSCCO lies just at the border)
at not too small fields ($a_0 \lesssim \lambda$). For such systems one
finds at $q_\perp = {q_\perp^*}$ and $q_z = {q_z^*}$:\cite{GK91}
\begin{mathletters} \label{c44.BSCCO}
\begin{eqnarray}
c_{44} &\approx& \frac{\Phi_0^2}{2 (4 \pi \gamma a_0 \lambda)^2}
\ln \frac{\gamma^2/\xi^2}{4 \pi \gamma^2/a_0^2 + \pi^2/d^2} ~,
\\
B_{cr} &\approx& \frac{\pi \Phi_0}{\gamma^2 d^2}
\ln(\gamma d/\xi) ~,
\end{eqnarray}
\end{mathletters}
and further on
\begin{mathletters} \label{wth.BSCCO}
\begin{eqnarray}
T_x& \approx & 
\left\{
\begin{array}{ll} 
c_L^2 \frac{d \Phi_0^2}{(4 \pi \lambda)^2}
& {\rm for } \quad B \gg B_{cr} \\
c_L^2 \frac{d \Phi_0^2}{(4 \pi \lambda)^2} 
\left( \frac{B_{cr}}{B} \right)^{1/2}
& {\rm for } \quad B \ll B_{cr} ~, \\
\end{array}
\right.
\\
T_z& \approx & 
\left\{
\begin{array}{ll} 
c_L^2 \frac{d \Phi_0^2}{(4 \pi \lambda)^2}
\frac{2}{\ln(B/B_{cr})}
& {\rm for } \quad B \gg B_{cr} \\
c_L^2 \frac{d \Phi_0^2}{(4 \pi \lambda)^2} 
\left( \frac{B_{cr}}{B} \right)
& {\rm for } \quad B \ll B_{cr} ~. \\
\end{array}
\right.
\end{eqnarray}
\end{mathletters}

The relative strength of fluctuations of in-plane- and $z$-bonds is
different for small and large fields. At large fields $w^{\rm
th}_x/a_0^2$ is field-independent, whereas $w^{\rm th}_z/a_0^2 \sim
1/\ln B$ increases without bounds, i.e. $w^{\rm th}_x \ll w^{\rm
th}_z$. Thus for increasing fields $T_x$ saturates at a constant
value, whereas $T_z$ vanishes. At small fields the situation is
reversed: $w^{\rm th}_x \gg w^{\rm th}_z$ because $w^{\rm th}_z/a_0^2
\sim B$ and $w^{\rm th}_x/a_0^2 \sim B^{1/2}$.

The bond widths provide more information than the typical displacement
$\langle {\bf u}^2 \rangle^{1/2}$ evaluated by the usual Lindemann
criterion. However, they are related by $\langle {\bf u}^2 \rangle
\approx \max(w^{\rm th}_x, w^{\rm th}_z)$. In our Lindemann criterion
(\ref{wth.BSCCO}) for the different types of phase transitions we use
the same Lindemann number $c_L$. In principle these values might be
different, since $z$-bonds and in-plane bonds are not not equivalent
from the point of view of lattice symmetries. But a use of different
Lindemann numbers would lead only to a quantitative shift of the phase
boundaries, which shall not be our main concern here.

In order to associate the bond widths to possible melting and
decoupling transitions, it is worth recalling the main features of
these transitions. Since for anisotropies $\gamma \gg 1$ the
interaction between vortices in the same layer is much weaker than the
interaction of vortices in different layers one might in principle
expect the possibility that melting occurs as a two-step process. In a
first step the positional correlations of vortices in different layers
could get lost while the correlations within the layers are preserved
(at least as quasi-long-range order). This remaining two-dimensional
order within the layers could then become short-ranged at a second
transition at a higher temperature. However, this scenario is probably
not realized, as one can conclude from an analogy to layered
$XY$-models. On symmetry grounds their phase diagram should be
topologically equivalent to that of layered crystals. It was
shown\cite{XY} that these models have only a single transition where
order is destroyed in all directions. Although the relative
fluctuations between the layers can become arbitrarily large for weak
coupling between the layers, which is the case for layered
superconductors at large fields, the transition temperature does not
drop below that of the two-dimensional system. In the Lindemann
approach melting has therefore to be identified with the breaking of
in-plane bonds.

The breaking of $z$-bonds therefore describes a different transition,
which is {\em not} related to the loss of crystalline order.  Two
possible transitions have been considered recently: the decoupling
transition and the supersolid transition. The decoupling transition of
layered superconductors\cite{GK91,DBMC93,KGL96} describes the loss of
phase coherence in the superconducting order parameter between
different layers and in general does not coincide with the melting
transition. The location of this transition can be estimated by a
Lindemann-type criterion for the phase difference between neighboring
layers. Although the vortex displacements are the main source of such
phase fluctuations, a Lindemann criterion for phase differences is in
general not equivalent to a Lindemann criterion for vortex
displacements, since a local vortex displacement leads to a nonlocal
phase distortion. However, at fields below the crossover field
$B_{cr}$, both types of Lindemann criteria actually are equivalent and
the decoupling transition is captured just by breaking of
$z$-bonds.\cite{GK91,DBMC93} At large fields the breaking of $z$-bonds
describes the location of the transition from a solid to a supersolid
vortex crystal due to a proliferation of vacancy and interstitial
lines in the vortex line crystal.\cite{FNF94}

\subsection{Pinning bond width}

After the brief review on the effect of thermal fluctuations on bond
widths and the possible interpretations of bond-breaking we now focus
on our main issue, the contribution of pinning to the bond widths and
the question how the above equilibrium phase diagram of the pure
evolves into a nonequilibrium phase diagram of the disordered system.

Only very recently the equilibrium phase diagram of the disordered
system ($v=0$) has received some attention (see
Refs. \CITE{KNH95,EN96,Vin,Kier96,dsF97,Giamarchi,KV97}).  In
particular the breaking of in-plane bonds has been used as indication
for the transition from a topologically ordered phase (Bragg-glass) to
a topologically disordered phase (vortex glass). For small magnetic
fields the quantitative evaluation of the bond widths turned out to be
quite a subtle issue because of the strong dispersion of the elastic
constants in layered materials and the breakdown of perturbation
theory in the low-field regime ($\xi \ll c_L a_0$). Since we are
eventually interested in the dynamic effects of disorder on the
dynamics, we do not attempt to reproduce the static results. However,
it is worthwhile to point out that in $D=3$ the disorder induced bond
fluctuations are {\em finite} for all bond types.  This finiteness is
found even at zero velocity and even if the bond widths are calculated
from Eq. (\ref{w.pin}) within naive perturbation theory that
overestimates the effect of disorder. For weak enough disorder all
bonds are stable and a topologically ordered phase can persist. For
low temperatures the main effect of weak disorder is to reduce the
long-range translational order to a quasi-long-range order.

At sufficiently large velocities the perturbative approach,
Eq. (\ref{w.pin}), can always be used to calculate the shift $\Delta
T_\alpha$ of the bond breaking temperatures $T_\alpha$ with respect to
the values $T^0_\alpha$ in the absence of disorder. For large
velocities these temperature shifts tend to zero and the use of the
lowest order perturbation theory is justified at least as long as the
disorder contributions to the bond widths satisfy $w^{\rm pin}
\lesssim \xi^2$.

The pinning bond widths  show a
complicated velocity dependence with several regimes, separated by
characteristic velocities related to the different elastic
constants. We focus here exclusively on the high-velocity regime $v
\gg c_{11}/a_0 \eta$ for which we estimate the pinning contribution
to the bond widths from Eq. (\ref{w.pin}):
\begin{mathletters} \label{w.pin.hiv}
\begin{eqnarray}
w_x^{\rm pin} &\approx&
\frac{a^2 \Delta_0}{\sqrt{8 \pi} a_0 d \xi^3 \eta^2 v^2} ~,
\\
w_y^{\rm pin} &\approx& 
\frac{a^2 \Delta_0}{4 \sqrt{2 \pi} \xi^3 d \eta v c_{11}} ~,
\\
w_z^{\rm pin} &\approx&
\frac{\pi a_0 \Delta_0}{16 \sqrt{2 \pi} \xi^3 \eta v
\sqrt{c_{11} c_{44}}} ~,
\end{eqnarray}
\end{mathletters}
where we used again $c_{11} {q_\perp^*}^2 \gg c_{44} {q_z^*}^2$.

From Eq. (\ref{w.pin.hiv}) one can immediately draw the important
conclusion that $y$- and $z$-bonds, for which $w^{\rm pin}$ decays
like $v^{-1}$, are affected by disorder at large velocities much
stronger than $x$-bonds with $w_x^{\rm pin} \sim v^{-2}$. As in $D=2$,
the relative strength of $y$-bond fluctuations in comparison to
$x$-bond fluctuations is in agreement with the prediction of a
smectic-like flow of the vortex lattice.\cite{BMR97,BMR97.hv}

In order to extract the net field dependence of the bond widths it is
important to remember the implicit dependences of the parameters. In
particular $\Delta_0 \sim a_0^{-4}$ since $\Delta$ is the correlator
of the pinning energy density in Eq. (\ref{Phi}). For the same reason
$\eta \sim a_0^{-2}$ and also all elastic constants carry a dominant
implicit field dependence $\sim a_0^{-2}$. Thus the relative
fluctuations $w_y^{\rm pin}/a_0^2 \sim a^0$ are essentially field
independent, whereas $w_x^{\rm pin}/a_0^2 \sim w_x^{\rm pin}/a_0^2
\sim a_0^{-1}$ increase for large fields. 

To visualize the velocity dependence of the dynamic transitions it is
instructive to translate the pinning bond widths (\ref{w.pin.hiv}) of
the driven vortex lattice into shifts $\Delta T_x$ of the
bond-breaking temperatures (\ref{wth.BSCCO}) of the pure system.  From
$w_\alpha ^{\rm th}+w_\alpha ^{\rm pin}= c_L^2 a_0^2$ the bond
breaking temperature is reduced by $\Delta T_\alpha = - T_\alpha
w_\alpha^{\rm pin}/c_L^2 a_0^2$. These shifts are
\begin{mathletters} \label{DT}
\begin{eqnarray}
\Delta T_x & \approx & -T_x \frac {\epsilon_0^2 \hat \Delta_0}
{4 \pi c_L^2 \xi^2 \eta_l^2 v^2 }\left(\frac{B}{H_{c2}}\right)^{1/2} ~,
\\
\Delta T_y & \approx &  -T_x \frac {\epsilon_0 \hat \Delta_0}
{4\sqrt{2 \pi} c_L^2 \xi \eta_l v } ~,
\\
\Delta T_z & \approx & -T_z \frac {\pi \epsilon_0 \hat \Delta_0}
{16\sqrt{2} c_L^2 \xi \eta_l v }\left(\frac{B}{B_{cr}}\right)^{1/2} ~.
\end{eqnarray}
\end{mathletters}
To make the net field dependence of bond widths explicit, we have
introduced a dimensionless disorder strength $\hat \Delta_0$ by
$\Delta_0=d (\xi \epsilon_0/a_0^2)^2 \hat \Delta_0$ and the friction
coefficient per line $\eta_l= a_0^2 \eta$. Furthermore we have used
again the values of the elastic constants for $q_\perp =
{q_\perp^*}$ and $q_z = {q_z^*}$, where $c_{11} \approx 4 c_{66}\approx
\epsilon_0 /a_0^2$, and $H_{c2}=\Phi_0 / 2 \pi \xi^2$.
In terms of these parameters the high-velocity regime is restricted by
$\eta_l v \gg \epsilon_0 / a_0$. 

In this high velocity regime $y$-bonds are less stable than $x$-bonds
over the whole field range. Dynamic melting is thus always driven by
the breaking of $y$-bonds.  In the absence of disorder the $z$-bonds
were more (less) stable than in-plane bonds for low (high) fields
compared to $B_{cr}$. This relation is preserved in the driven case,
since the bond breaking temperature of in-plane bonds is shifted more
(less) than that of the $z$-bonds at low (high) fields. Thus, the
topology of the phase diagram of the driven vortex lattice remains
unchanged at large velocity, as one could expect naively. The
high-velocity expansion does not provide evidence for a dynamic shift
of the crossover field.

\section{Discussion and conclusions}
\label{sec.conc}

To discuss the actual observability of the high-velocity regime it is
important to compare the above velocity range to the depairing
velocity of Cooper pairs. From the depairing current (see
e.g. Ref. \CITE{blat94}) the depairing velocity can be estimated as
$\eta_l v_{\rm depair} \sim \epsilon_0/\xi$. Thus even if $\xi$ is not
much smaller than $a_0$ (as required by the validity of the
perturbative approach) the high-velocity regime covers a substantial
current range below the depairing current and should be experimentally
accessible. As long as disorder is weak ($\hat \Delta_0 \ll 1$) the
location of the dynamic melting and decoupling transitions is close to
the location of the equilibrium transition of the pure system,
i.e. $\Delta T_\alpha \ll T_\alpha$.

In this high-velocity regime dynamic melting is always driven by the
breaking of $y$-bonds. The dynamic shifts of both melting and
decoupling transition temperatures scale like $\Delta T_\alpha \sim
v^{-1}$. This dependence is in agreement with experiments\cite{kap}
and numerical simulations\cite{AKV97,OSZ97} and the original concept
of the shaking temperature.\cite{KV94} We have extended this original
approach, which was focused on the dynamic response of a single vortex
and on the response of the lattice at large scales, on the
fluctuations of bonds. Thereby we were able to put the
characterization of the nonequilibrium transitions on a common basis
with the Lindemann approaches to the equilibrium counterparts.

At smaller drive the velocity dependence of the bond widths and the
temperature shifts will be in general weaker than at high drive. In two
dimensions the velocity dependence becomes logarithmic at small
velocities and the crossover between different regimes has been
calculated numerically (see Fig. \ref{fig_wpin}). In three
dimensions there are even more dynamic regimes between 
the asymptotic regimes of zero and large velocity because of the complex
elastic properties of the vortex lattice.

However, from Eq. (\ref{DT}) we can conclude that for weak disorder
($\hat \Delta_0 \ll 1$) the scaling of the high-velocity regime should
be observable in a considerable velocity range below depairing. In
this case, where the temperature shifts are small ($\Delta T_\alpha
\ll T_\alpha$), it should be possible to observe a dynamic shift of
the phase boundaries experimentally. To the best of our knowledge,
such dynamic shifts have not been reported. We naturally expect a
dynamic shift to be most pronounced in the vicinity of the critical
point of the melting line since the main effect of a driving current
is to reduce the effective strength of disorder and this point has
been found to be quite sensitive to the strength of disorder in
equilibrium .\cite{Khay+97} In particular the location of the critical
point will be shifted to larger fields by a driving current.

The anisotropy of bond fluctuations found above is consistent with the
picture of smectic vortex flow\cite{BMR97,BMR97.hv,SV97.hv} due to the
dominant proliferation of dislocations with Burger's vectors parallel
to the drift velocity, providing the actual mechanism for bond
breaking. We expect the anisotropy of bond fluctuations to have
immediate consequences for the structure of the coherently flowing
vortex lattice. In particular even the Bravais basis of the vortex
lattice will display this anisotropy. In the presence of disorder the
vortex system will show an expansion of its lattice constants similar
to a thermal expansion. From the anisotropy of the bond fluctuations
one has to expect that this expansion will be larger in $y$-direction
than in $x$-direction. It would be interesting to analyze neutron
scattering and Bitter decoration experiments on the driven vortex
lattice in view of a dynamic shift of the location of the
(quasi)-Bragg peaks.

\section*{Acknowledgments}

The authors gratefully acknowledge useful discussions with 
J. Kierfeld, H. Nordborg, and in particular with A.E. Koshelev. 

This work was supported from Argonne National Laboratory through the
U.S.  Department of Energy, BES-Material Sciences, under contract No.
W-31-109-ENG-38 and by the NSF-Office of Science and Technology
Centers under contract No.  DMR91-20000 Science and Technology Center
for Superconductivity. S. acknowledges support by the Deutsche
Forschungsgemeinschaft project SFB341 and grant SCHE/513/2-1.

\newpage

\end{multicols}
\end{document}